\documentclass[submission,copyright,creativecommons]{eptcs}
\providecommand{\event}{FESCA 2013}

\usepackage{breakurl}             
\usepackage{amsmath}
\usepackage{amssymb}
\usepackage{amsthm}
\usepackage{amscd}
\usepackage{graphicx}
\usepackage[cmyk]{xcolor}
\usepackage{lscape}
\usepackage{proof}
\usepackage{tikz}
\usetikzlibrary{arrows}
\usepackage[all]{xy}
\usepackage{verbatim}
\usepackage{bm}
\usepackage{url}

\title{Assembling the Proofs of\\Ordered Model Transformations}

\author{Maribel Fern\'{a}ndez \quad\quad Jeffrey Terrell
\institute{
Department of Informatics,
King's College London,
Strand,
London WC2R 2LS,
UK\\
\email{jeffrey.terrell@kcl.ac.uk}}
}

\def\titlerunning{Ordered Model Transformations}
\def\authorrunning{M. Fern\'{a}ndez \& J. Terrell}

\newtheorem{definition}{Definition}
\newtheorem{property}{Property}

\begin{document}
\maketitle

\begin{abstract}
In model-driven development, an ordered model transformation is a nested set of transformations between source and target classes, in which each transformation is governed by its own pre and postconditions, but structurally dependent on its parent. Following the proofs-as-model-transformations approach, in this paper we consider a formalisation in Constructive Type Theory of the concepts of model and model transformation, and show how the correctness proofs of potentially large ordered model transformations can be systematically assembled from the proofs of the specifications of their parts, making them easier to derive.
\end{abstract}


\newcommand{\nil}{[\,]}
\newcommand{\df}{\:=_{df}\:}
\newcommand{\dftwo}{=_{df}}
\newcommand{\doubleplus}{+\negthickspace+\medspace}
\newcommand{\twoheadleftrightarrow}{\twoheadleftarrow \hspace{-0.2cm} \twoheadrightarrow}
\newcommand{\eg}{e.g.\ }
\newcommand{\ie}{i.e.\ }
\newcommand{\combinator}[1]{\texttt{#1}}
\newcommand{\parentobject}[1]{{#1}-}
\newcommand{\childobject}[1]{{#1}+}
\newcommand{\objectlist}[2]{\vec{#1}_{#2}}
\newcommand{\leafsubscript}{\;\dashv}
\newcommand{\cond}{\Rightarrow}
\newcommand{\bicond}{\Leftrightarrow}
\newcommand{\domain}[1]{\vert \mathcal{#1} \vert}
\newcommand{\interpretation}[1]{\mathcal{#1}}
\newcommand{\makestrue}[2]{\interpretation{#1} \vDash {#2}}
\newcommand{\xdotx}[2]{{#1}, \dots, {#2}}
\newcommand{\mr}[2]{{#1}\; \underline{\text{mr}}\; {#2}}
\newcommand{\product}[2]{{{#1} \times {#2}}}


\newcommand{\ptype}[1]{{#1}\ \text{is a type}}
\newcommand{\pfrule}[1]{({#1}\,F)}
\newcommand{\pirule}[2]{({#1}\,I_{#2})}
\newcommand{\perule}[2]{({#1}\,E_{#2})}
\newcommand{\pencode}[1]{\varepsilon_{#1}}

\newcommand{\pimplies}{\rightarrow}
\newcommand{\piff}{\leftrightarrow}
\newcommand{\pmetaimplies}{\Rightarrow}
\newcommand{\pdeduce}[2]{{#1} \vdash {#2}}
\newcommand{\passumption}[2]{[{#2}]^{\,#1}}
\newcommand{\pincontext}[1]{\ulcorner{#1}\urcorner}
\newcommand{\penviron}[1]{\ulcorner{#1}\urcorner}
\newcommand{\passrule}{(Ass)}
\newcommand{\pobject}[2]{#1 \colon {#2}}

\newcommand{\parrowirule}[1]{(\pimplies\!I)_{\,#1}}
\newcommand{\parrowerule}{(\pimplies\!E)}
\newcommand{\parrowfrule}{(\pimplies\!F)}
\newcommand{\pnirule}[1]{(NI)_{#1}}
\newcommand{\pnatirule}[1]{(\pnat\,I_{#1})}
\newcommand{\pandirule}{(\land\,I)}
\newcommand{\panderule}[1]{(\land\,E_{#1})}
\newcommand{\porirule}[1]{(\lor I_{#1})}
\newcommand{\porerule}{(\lor E)}
\newcommand{\pexistsirule}{(\exists\,I)}
\newcommand{\pexistserule}[1]{(\exists\,E_{#1})}
\newcommand{\pexistsfrule}{(\exists\,F)}
\newcommand{\pexistseprimerule}{(\exists\,E')}
\newcommand{\pexistsenoprimerule}{(\exists\,E)}
\newcommand{\pforallirule}[1]{(\forall I)_{#1}}
\newcommand{\pforallerule}{(\forall E)}
\newcommand{\pforallfrule}{(\forall F)}
\newcommand{\piirule}{(II)}
\newcommand{\pierule}{(IE)}
\newcommand{\piiprimerule}{(II')}
\newcommand{\plisterule}{(list\ E)}
\newcommand{\pabsurderule}{(\perp\negmedspace E)}
\newcommand{\pboolerule}{(Bool\;E)}
\newcommand{\pboolirule}[1]{(Bool\;I_{#1})}
\newcommand{\pequaldefrule}{(\df)}
\newcommand{\pdefrule}{(\df)}
\newcommand{\pconvertrule}{(=_{red})}
\newcommand{\predrule}{(=_{red})}
\newcommand{\pabsrule}{(=_{abs})}
\newcommand{\pextractrule}{(=_{extract})}
\newcommand{\pmrrule}{(=_{mr})}
\newcommand{\pequalorrule}{(=_{or})}
\newcommand{\pnotequalorrule}{(\neq_{or})}
\newcommand{\ptauirule}{(\tau I)}

\newcommand{\ppair}[2]{\langle #1,\, #2 \rangle}
\newcommand{\ptriple}[3]{\langle #1,\, #2,\, #3 \rangle}
\newcommand{\pfirst}[1]{f\!st\,#1}
\newcommand{\psecond}[1]{snd\,#1}
\newcommand{\pthird}[1]{thd\,#1}
\newcommand{\plast}[1]{lst\,#1}
\newcommand{\psource}[1]{{#1}_{src}}
\newcommand{\ptarget}[1]{{#1}_{tgt}}
\newcommand{\ppi}[2]{\pi_{#1}\,#2}
\newcommand{\pstop}{\;.}
\newcommand{\pcomma}{\;,}
\newcommand{\psemicolon}{\;;}
\newcommand{\pleft}[1]{inl\;#1}
\newcommand{\pright}[1]{inr\;#1}
\newcommand{\pcases}[3]{cases\;{#1}\;{#2}\;{#3}}
\newcommand{\plistrec}[3]{lrec\;{#1}\;{#2}\;{#3}}
\newcommand{\preduction}[1]{\rightarrow_{#1}}
\newcommand{\papplication}[2]{({#1}\:{#2})}
\newcommand{\papp}[2]{({#1}\:{#2})}
\newcommand{\psmallapplication}[2]{{#1}\:{#2}}
\newcommand{\pappnp}[2]{{#1}\:{#2}}
\newcommand{\pappthree}[3]{\papp{#1}{\papp{#2}{#3}}}
\newcommand{\pappfour}[4]{\papp{#1}{\papp{#2}{\papp{#3}{#4}}}}
\newcommand{\pappnpthree}[3]{\pappnp{#1}{\pappnp{#2}{#3}}}
\newcommand{\pappnpfour}[4]{\pappnp{#1}{\pappnp{#2}{\pappnp{#3}{#4}}}}
\newcommand{\pappnpfive}[5]{\pappnp{#1}{\pappnp{#2}{\pappnp{#3}{\pappnp{#4}{#5}}}}}
\newcommand{\pappnpsix}[6]{\pappnp{#1}{\pappnp{#2}{\pappnp{#3}{\pappnp{#4}{\pappnp{#5}{#6}}}}}}
\newcommand{\pappnpseven}[7]{\pappnp{#1}{\pappnp{#2}{\pappnp{#3}{\pappnp{#4}{\pappnp{#5}{\pappnp{#6}{#7}}}}}}}
\newcommand{\pappnpeight}[8]{\pappnp{#1}{\pappnp{#2}{\pappnp{#3}{\pappnp{#4}{\pappnp{#5}{\pappnp{#6}{\pappnp{#7}{#8}}}}}}}}
\newcommand{\pappnpnine}[9]{\pappnp{#1}{\pappnp{#2}{\pappnp{#3}{\pappnp{#4}{\pappnp{#5}{\pappnp{#6}{\pappnp{#7}{\pappnp{#8}{#9}}}}}}}}}
\newcommand{\psubstitute}[3]{{#1}\,[{#2}/{#3}]}
\newcommand{\pwitness}[1]{\overline{#1}}
\newcommand{\panonwitness}[1]{\hat{#1}}
\newcommand{\pexists}[3]{\exists \pobject{#1}{#2}\,.\,{#3}}
\newcommand{\pforall}[3]{\forall \pobject{#1}{#2}\,.\,{#3}}

\newcommand{\qforall}[2]{\forall {#1}\,.\,{#2}}
\newcommand{\qforalltwo}[3]{\qforall{{#1}}{\qforall{{#2}}{{#3}}}}
\newcommand{\qforallthree}[4]{\qforall{{#1}}{\qforall{{#2}}{\qforall{{#3}}{{#4}}}}}
\newcommand{\qforallfour}[5]{\qforall{{#1}}{\qforall{{#2}}{\qforall{{#3}}{{\qforall{{#4}}{#5}}}}}}
\newcommand{\qforallfive}[6]{\qforall{{#1}}{\qforall{{#2}}{\qforall{{#3}}{{\qforall{{#4}}{\qforall{{#5}}{#6}}}}}}}
\newcommand{\qforallsix}[7]{\qforall{{#1}}{\qforall{{#2}}{\qforall{{#3}}{{\qforall{{#4}}{\qforall{{#5}}{\qforall{{#6}}{#7}}}}}}}}
\newcommand{\qforallseven}[8]{\qforall{{#1}}{\qforall{{#2}}{\qforall{{#3}}{{\qforall{{#4}}{\qforall{{#5}}{\qforall{{#6}}{\qforall{{#7}}{#8}}}}}}}}}
\newcommand{\qforalleight}[9]{\qforall{{#1}}{\qforall{{#2}}{\qforall{{#3}}{{\qforall{{#4}}{\qforall{{#5}}{\qforall{{#6}}{\qforall{{#7}}{\qforall{{#8}}{#9}}}}}}}}}}

\newcommand{\plambda}[2]{\lambda {#1}\,.\,{#2}}
\newcommand{\plambdatwo}[3]{\plambda{{#1}}{\plambda{{#2}}{{#3}}}}
\newcommand{\plambdathree}[4]{\plambda{{#1}}{\plambda{{#2}}{\plambda{{#3}}{{#4}}}}}
\newcommand{\plambdafour}[5]{\plambda{{#1}}{\plambda{{#2}}{\plambda{{#3}}{{\plambda{{#4}}{#5}}}}}}
\newcommand{\plambdafive}[6]{\plambda{{#1}}{\plambda{{#2}}{\plambda{{#3}}{{\plambda{{#4}}{\plambda{{#5}}{#6}}}}}}}
\newcommand{\plambdasix}[7]{\plambda{{#1}}{\plambda{{#2}}{\plambda{{#3}}{{\plambda{{#4}}{\plambda{{#5}}{\plambda{{#6}}{#7}}}}}}}}
\newcommand{\plambdaseven}[8]{\plambda{{#1}}{\plambda{{#2}}{\plambda{{#3}}{{\plambda{{#4}}{\plambda{{#5}}{\plambda{{#6}}{\plambda{{#7}}{#8}}}}}}}}}
\newcommand{\plambdaeight}[9]{\plambda{{#1}}{\plambda{{#2}}{\plambda{{#3}}{{\plambda{{#4}}{\plambda{{#5}}{\plambda{{#6}}{\plambda{{#7}}{\plambda{{#8}}{#9}}}}}}}}}}

\newcommand{\pifthenelse}[3]{i\!f\;{#1}\;then\;{#2}\;else\;{#3}}
\newcommand{\pifthenelseshort}[3]{\pappnp{i\!f}{#1}{#2}{#3}}
\newcommand{\pconvert}[2]{{#1} \twoheadleftrightarrow {#2}}
\newcommand{\pfrom}[1]{\eqref{#1}}
\newcommand{\ppath}[3]{\widehat{#1}_{#2, #3}}
\newcommand{\prootpath}[2]{\widehat{#1}_{#2}}
\newcommand{\realize}{\rightsquigarrow}
\newcommand{\prealize}[2]{{#1} \realize {#2}}
\newcommand{\preducesto}{\twoheadrightarrow}
\newcommand{\pbetareduction}{\twoheadrightarrow_{\beta}}
\newcommand{\pbetacon}{\twoheadrightarrow_{\beta}}
\newcommand{\palphacon}{\rightarrow_{\alpha}}
\newcommand{\pcomputeone}{\rightarrow}
\newcommand{\pcomputemany}{\twoheadrightarrow}
\newcommand{\pcomputesto}{\twoheadrightarrow}
\newcommand{\pecases}[4]{ecases_{{#1}, {#2}}\;{#3}\;{#4}}
\newcommand{\pvcases}[5]{vcases_{{#1}, {#2}}\;{#3}\;{#4}\;{#5}}
\newcommand{\pposthighlight}{\;^*}
\newcommand{\pprehighlight}{^*\;}
\newcommand{\prealizer}[2]{\phi({#2})^{#1}}
\newcommand{\pabort}[2]{abort_{#1}\;{#2}}
\newcommand{\pmyctt}{TT_0}
\newcommand{\pmrnull}{\epsilon}
\newcommand{\pcatjudge}[1]{\vdash {#1}}
\newcommand{\phypjudge}[2]{{#1} \vdash {#2}}
\newcommand{\pjudge}[2]{{#1} \vdash {#2}}
\newcommand{\pjudgefst}[1]{\mathfrak{F}_{#1}}
\newcommand{\pjudgesnd}[1]{\mathfrak{S}_{#1}}\newcommand{\pproof}[1]{Proof({#1})}\newcommand{\pglobals}{\mathcal{E}}
\newcommand{\plist}[1]{[\,{#1}\,]}

\newcommand{\pcontext}[1]{[{#1}]}
\newcommand{\pcontextditto}{[\dots]}
\newcommand{\pcontexttwo}[2]{[{#1},\,{#2}]}
\newcommand{\pcontextthree}[3]{[{#1},\,{#2},\,{#3}]}
\newcommand{\pcontextfour}[4]{[{#1},\,{#2},\,{#3},\,{#4}]}
\newcommand{\pcontextfive}[5]{[{#1},\,{#2},\,{#3},\,{#4},\,{#5}]}
\newcommand{\pcontextsix}[6]{[{#1},\,{#2},\,{#3},\,{#4},\,{#5},\,{#6}]}
\newcommand{\pcontextseven}[7]{[{#1},\,{#2},\,{#3},\,{#4},\,{#5},\,{#6},\,{#7}]}
\newcommand{\pcontexteight}[8]{[{#1},\,{#2},\,{#3},\,{#4},\,{#5},\,{#6},\,{#7},\,{#8}]}
\newcommand{\plistnine}[9]{[\,{#1},\,{#2},\,{#3},\,{#4},\,{#5},\,{#6},\,{#7},\,{#8},\,{#9}]}

\newcommand{\pemptylist}{[\,]}
\newcommand{\pset}[1]{\lbrace {#1} \rbrace}
\newcommand{\pemptyset}{\lbrace\,\rbrace}
\newcommand{\ptuple}[1]{({#1})}
\newcommand{\ptwotuple}[2]{({#1},\;{#2})}
\newcommand{\ppifun}[2]{\pi_{#1}^{#2}}
\newcommand{\pinhab}[1]{\langle {#1} \rangle}
\newcommand{\punspecified}{\Box}
\newcommand{\poption}[1]{opt\; {#1}}
\newcommand{\psome}[1]{some\, {#1}}
\newcommand{\pnone}{none}
\newcommand{\ppartorder}[3]{({#1}\, {#2}\, {#3})}
\newcommand{\pbuild}[1]{@_{#1}}
\newcommand{\pconstruct}[1]{@_{#1}}
\newcommand{\pdestruct}[1]{@^{-1}_{#1}}
\newcommand{\pbar}{\; | \;}
\newcommand{\pid}[1]{Id_{#1}}
\newcommand{\pnat}{Nat}
\newcommand{\pmatchnat}[5]{match_{\pnat}({#1}, \ppair{#2}{#3}, \ppair{#4}{#5})}
\newcommand{\pmatchbool}[5]{match_{\pbool}({#1}, \ppair{#2}{#3}, \ppair{#4}{#5})}
\newcommand{\pmatchone}[4]{match_{#1}({#2}, \ppair{#3}{#4})}
\newcommand{\pmatchtwo}[6]{match_{#1}({#2}, \ppair{#3}{#4}, \ppair{#5}{#6})}
\newcommand{\pnatzero}{O}
\newcommand{\pnatsucc}[1]{\pappnp{S}{#1}}
\newcommand{\pnatone}{\pnatsucc{\pnatzero}}
\newcommand{\pnattwo}{\pnatsucc{\pnatone}}
\newcommand{\pnatthree}{\pnatsucc{\pnattwo}}
\newcommand{\preflex}[1]{r({#1})}
\newcommand{\ppropuni}{Prop}
\newcommand{\psetuni}{Set}
\newcommand{\ptypeuni}{Type}
\newcommand{\psmalluni}{U}
\newcommand{\puniverse}[1]{U_{#1}}
\newcommand{\ppre}[1]{Pre_{#1}}
\newcommand{\ppost}[1]{Post_{#1}}
\newcommand{\pland}{\: \land}
\newcommand{\plor}{\: \lor}
\newcommand{\pwild}{\star}
\newcommand{\pbooltrue}{True}
\newcommand{\pboolfalse}{False}
\newcommand{\pbool}{Bool}
\newcommand{\pbaseat}[1]{Id_{#1}}
\newcommand{\prefat}[1]{R_{#1}}
\newcommand{\pcomment}[1]{}
\newcommand{\pcert}[1]{\surd {#1}\,}

\newcommand{\pplambda}[2]{\lambda {#1}.{#2}}
\newcommand{\ppobject}[2]{#1 \colon {#2}}
\newcommand{\pgoal}[1]{Goal\; #1}

\newcommand{\predefaz}{
\def\pA{A} \def\pB{B} \def\pC{C} \def\pD{D} \def\pE{E} \def\pF{F} 
\def\pG{G} \def\pH{H} \def\pI{I} \def\pJ{J} \def\pK{K} \def\pL{L} 
\def\pM{M} \def\pN{N} \def\pO{O} \def\pP{P} \def\pQ{Q} \def\pR{R} 
\def\pS{S} \def\pT{T} \def\pU{U} \def\pV{V} \def\pW{W} \def\pX{X} 
\def\pY{Y} \def\pZ{Z}
\def\pAA{AA} \def\pBB{BB} \def\pCC{CC} \def\pDD{DD} \def\pEE{EE} \def\pFF{FF} 
\def\pGG{GG} \def\pHH{HH} \def\pII{II} \def\pJJ{JJ} \def\pKK{KK} \def\pLL{LL} 
\def\pMM{MM} \def\pNN{NN} \def\pOO{OO} \def\pPP{PP} \def\pQQ{QQ} \def\pRR{RR} 
\def\pSS{SS} \def\pTT{TT} \def\pUU{UU} \def\pVV{VV} \def\pWW{WW} \def\pXX{XX} 
\def\pYY{YY} \def\pZZ{ZZ}
}


\newcommand{\pdschoice}[3]{{#1} \pimplies {#2}\; []\; {#3}}


\newcommand{\umlclass}[1]{\texttt{#1}}
\newcommand{\umlattribute}[1]{\texttt{#1}}
\newcommand{\umlrelationship}[1]{\texttt{R{#1}}}


\newcommand{\pmodel}[1]{{#1}}
\newcommand{\ptransformation}[2]{{#1} \mapsto {#2}}
\newcommand{\prung}[2]{{#1} \mapsto {#2}}


\newcommand{\coqobject}[2]{\coq{{#1}:\;{#2}}}
\newcommand{\coqforall}[3]{\coq{forall \coqobject{#1}{#2}, {#3}}}
\newcommand{\coqexists}[3]{\coq{exists \coqobject{#1}{#2}, {#3}}}
\newcommand{\coqfun}[3]{\coq{fun \coqobject{#1}{#2} => {#3}}}
\newcommand{\coqfunapp}[2]{\coq{{#1} {#2}}}
\newcommand{\coqand}[2]{\coq{{#1} \coqandsymbol\ {#2}}}
\newcommand{\coqor}[2]{\coq{{#1} \coqorsymbol\ {#2}}}
\newcommand{\coqimply}[2]{\coq{{#1} \coqimplysymbol\ {#2}}}
\newcommand{\coqandsymbol}{/\symbol{92}}
\newcommand{\coqorsymbol}{\symbol{92}/}
\newcommand{\coqimplysymbol}{->}
\newcommand{\coqtactic}[1]{\coq{#1}}


\newcommand{\prologdef}[1]{\texttt{#1}}
\newcommand{\prolog}[1]{\textcolor{violet}{\prologdef{#1}}}
\newcommand{\prologalt}[1]{\textcolor{red}{\prologdef{#1}}}


\newcommand{\pcode}[1]{\texttt{#1}}
\newcommand{\codedef}[1]{\texttt{#1}}
\newcommand{\code}[1]{\textcolor{blue}{\prologdef{#1}}}
\newcommand{\codealt}[1]{\textcolor{green}{\prologdef{#1}}}


\newcommand{\pcolor}[2]{\color{#1} {#2} \color{black}} 
\newcommand{\pconfig}[3]{\mathbf{\langle \pcolor{blue}{#1},\ \pcolor{red}{#2},\ \pcolor{violet}{#3} \rangle}}
\newcommand{\pconstructor}[1]{\color{red} {#1} \color{black}}
\newcommand{\pwhiledo}[2]{while\ {#1}\ do\ {#2}}
\newcommand{\ptransition}[6]{\pconfig{#1}{#2}{#3} \rightarrow \pconfig{#4}{#5}{#6}}
\newcommand{\pbigsteppair}[2]{\ppair{\pcolor{blue}{#1}}{\pcolor{red}{#2}}}
\newcommand{\pbigstep}[4]{\pbigsteppair{#1}{#2} \Downarrow \pbigsteppair{#3}{#4}}
\newcommand{\psmallsteppair}[2]{\ppair{\pcolor{blue}{#1}}{\pcolor{red}{#2}}}
\newcommand{\psmallstep}[4]{\psmallsteppair{#1}{#2} \rightarrow \psmallsteppair{#3}{#4}}

\section{Introduction}
\label{sec:1-i}

In this paper, we outline a mechanism to assemble correctness proofs
of model transformations in the context of Model Driven Development
(MDD). Although MDD is in widespread use, it is essentially an informal approach to software development which does not
guarantee the correctness of model transformations. High-trust
solutions are essential if MDD is to be used in safety critical systems and beyond.

The problem of establishing the correctness of a model transformation
is well established, and work has been done towards formalising the
process using for instance rewriting languages (e.g. Maude~\cite{ClavelM:maude}) or
typed multigraphs~\cite{Thirioux}. However, these approaches are
first-order and do not reflect the higher-order nature of the
UML-based techniques.  The aim of our research is to lay the
foundations on which a range of \emph{certified} model transformations
might be built, following a line of work that started
in~\cite{poernomo-2008}, where the use of constructive type theory to
implement model transformations was first discussed. The notion of an ordered model transformation was introduced in~\cite{poernomo-2010}, to describe how a complex transformation between models, built from a potentially large number of interrelated classes, might be derived from the specification of a series of mappings between classes, via a partially ordered traversal of the source and target models. This paper represents a significant advance on that work in that it a) formally defines the specification of an ordered model transformation in type theory, and b) provides a mechanism for assembling the proofs of ordered model transformations from their constituent parts.

In this paper, a \emph{model} is a Unified Modelling Language (UML)~\cite{omg-uml-2009} class model, and a
\emph{model transformation} is a function which maps the artefacts of a
source model (classes, attributes and relationships) onto the artefacts of a
target model~\cite{lano-2009-2,lano-2009-1}. UML is a graphical language for specifying the structure and behaviour of object oriented systems. It is also a pillar of the Object Management Group's (OMG) Model Driven Architecture (MDA) \cite{omg-mda-2003} (a particular brand of MDD), along with the transformation language Query/View/Transform (QVT)~\cite{omg-qvt-2011} . 

Consider a
transformation between two models (see Fig. \ref{fig:1-i.red})
\begin{figure}
\centerline{\xymatrix{
\ar@{->}[d]^{R}_>>{*} && 
\ar@{->}[d]^{S}_>>{*}\\
X
\ar@{->}[d]^{R'}_>>{*} 
\ar@{|-->}[rr]^<<<<{Pre_X}^>>>>{Post_Y} && 
Y
\ar@{->}[d]^{S'}_>>{*}\\
&&
}}
\caption{A transformation between classes $X$ and $Y$, which is subject to a precondition on $X$ and a postcondition on $X$ and $Y$.}
\label{fig:1-i.red}
\end{figure}
in which each object of $X$ is transformed into an object of $Y$, via a precondition at $X$ and a postcondition at $Y$. In general, the postcondition at $Y$ is composed of three components: $Data$ asserts a relation between the attributes of $X$ and $Y$; $Link$ asserts a relation between $Y$ and the class that contains it; and $Nest$ defines the specification (in context) of the transformation between the classes that $X$ and $Y$ contain (clearly, if $Y$ were a root class, the postcondition would not have a $Link$ component, and if $Y$ were a leaf class, the postcondition would not have a $Nest$ component).

While $Data$ components vary significantly between transformations (there is no reason why they should be the same), the $Link$ and $Nest$ components are generally quite similar. In fact the only assertion that a $Link$ component can make is that $Y$ participates in a relationship with the class that contains it; and all that a $Nest$ component can do is pass control of the specification to the classes that $X$ and $Y$ contain. This opens up the prospect of removing from users the tedious task of proving $Link$ components by hand.\footnote{The proof of an assertion involving a many-valued relationship requires a proof by list induction, and the proof of a chain of many-valued relationships, which is not uncommon, requires a nested set of proofs by list induction.} Of course, this prospect only presents itself by virtue of the \emph{ordered} nature of the models and transformations under consideration, where order is defined by \emph{containment}. However, such transformations are sufficiently common in practice (see, for instance,~\cite{lano-2009-2} for examples) to make this a worthwhile pursuit.

In this paper, we focus on a particular but nonetheless ubiquitous
kind of model transformation, in which the source and target models
are either partially or totally ordered.\footnote{Hierarchical models
  are the rule rather than the exception in industry (the UML
  metamodel is fundamentally hierarchical), where transformations are
  notable for their size rather than their complexity.}  In
particular, based on the definitions of model and model transformation
given in~\cite{poernomo-2008}, we show that the proofs of the
specifications of large ordered model transformations can be
systematically assembled from their parts, making them easier to
derive. 
Our main contribution is a method to derive correctness proofs for
ordered model transformations by assembling the proofs of their parts,
within constructive type theory. We illustrate the method with examples.

The proofs in this paper have all been implemented in the Coq Proof Assistant~\cite{bertot-2004}, see the Coq scripts at \url{http://www.inf.kcl.ac.uk/pg/terrellj}.

The paper is organised as follows:
In Section \ref{sec:2-p}, we give a brief introduction to MDA and type theory, to try to make the paper self contained. 
In Section~\ref{sec:3-ttfmt}, we show how to formalise a model transformation (the specification and its correctness proof)  in constructive type theory, including the key notion of a parametric proof (a proof with a  \textit{hole} over which it is possible to quantify and hence parametrise). We then use this idea to formally specify ordered model transformations in general in Section~\ref{sec:4-omt}. In Section~\ref{sec:A-ce}, there is a concrete example of an assembled proof. Finally, in Section \ref{sec:5-c}, we sum up and discuss future developments.

\section{Preliminaries}
\label{sec:2-p}

We recall the basic notions of Model Driven Architecture (MDA) and Constructive Type Theory (CTT) that are used in the paper. We refer the reader to~\cite{lano-2009-2} and~\cite{martin-lof-1984} for more details on MDA and CTT, respectively.

\subsection{Model Driven Architecture}

The movement away from the machine to a higher level of abstraction began in earnest in the early 1990s, with the advent of a number of object-oriented analysis and design methodologies. The most influential, in the authors' view, was the one proposed by Shlaer and Mellor in \cite{shlaer-mellor-1988, shlaer-mellor-1991}, for it played a huge part in shaping the MDA a decade later. The aims of the MDA are two-fold: first, that software systems should be developed independently of the platforms on which they will eventually run, and second, that they should be translated into specific implementations using standard parts, namely model to text and model to model transformations.


In its simplest form, a model transformation takes as input a model, written in a source modelling language (IL), and outputs a new model, written in a possibly different target modelling language (OL). The transformation should be applicable to any model written in IL, therefore it can be seen as a mapping from elements of IL to elements of OL.

In MDA, both the input and output languages are defined as metamodels within the Meta-Object Facility (MOF) \cite{omg-mof-2006}. Metamodelling in the MOF is usually done according to a four level hierarchy~\cite{mellor-2004}.
The levels are related by an object-oriented style class/object instantiation relationship: classes at level $M_{i+1}$ provide descriptions of objects at level $M_i$. Roughly speaking, we can think of entities at the $M_0$ level as objects representing instances of an $M_1$ UML class. The $M_2$ level is where metamodels are defined. Metamodels are collections of instances of the $M_3$ level classes (meta-meta-classes). The $M_3$ level of the MOF model is used to classify the elements that make up an $M_2$ level metamodel. 

Following~\cite{poernomo-2008,poernomo-2010}, in this paper we will consider model transformations as higher-order functional programs satisfying certain pre and post conditions. 

\subsection{Constructive Type Theory}
The type theory below is based on the one proposed by Martin-L\"{o}f \cite{martin-lof-1984}.
A type is defined by prescribing how its inhabitants are formed. For example, if $S$ is the successor function, then the inhabitants of the type $\pnat$ are given by
\[
\begin{matrix}
\infer[]{\pobject{0}{\pnat}}
{
}
& \hspace{1cm}
\infer[\pstop]{\pobject{\papp{S}{n}}{\pnat}}
{
	\pobject{n}{\pnat}
}
\end{matrix}
\]

If $A$ and $B$ are types, then $A \land B$, $A \lor B$ and $A \pimplies B$ are defined to be types too, where $A \land B$ is inhabited by a pair of inhabitants of $A$ and $B$, $A \lor B$ is inhabited by an inhabitant of $A$ or $B$, together with an indication as to whether it is an inhabitant of $A$ (on the left) or $B$ (on the right), and $A \pimplies B$ is inhabited by a function from $A$ to $B$, i.e.
\[
\begin{matrix}
\infer[]{\pobject{\ppair{a}{b}}{A \land B}}
{
	\pobject{a}{A}
	&
	\pobject{b}{B}
}
& \hspace{.5cm}
\infer[]{\pobject{\pleft{a}}{A \lor B}}
{
	\pobject{a}{A}
}
& \hspace{.5cm}
\infer[]{\pobject{\pright{b}}{A \lor B}}
{
	\pobject{b}{B}
}
& \hspace{.5cm}
\infer[\pstop]{\pobject{\plambda{\pobject{a}{A}}{b}}{A \pimplies B}}
{
	\infer*[]{\pobject{b}{B}}
	{
		\passumption{}{\pobject{a}{A}}
	}
}
\end{matrix}
\]
If $a$ is an inhabitant of $A$, and $B(a)$ is a type whose inhabitants depend on $a$, then $\pforall{a}{A}{B(a)}$ and $\pexists{a}{A}{B(a)}$ are defined to be types, where $\pforall{a}{A}{B(a)}$ is inhabited by a function that takes $A$ to $B(a)$, and $\pexists{a}{A}{B(a)}$ is inhabited by a pair of inhabitants of $A$ and $B(a)$, i.e. 
\[\begin{matrix}
\infer[]{\pobject{
\plambda{\pobject{a}{A}}{b}
}{
\pforall{a}{A}{B(a)}
}}
{
	\infer*[]{\pobject{b}{B(a)}}
	{
		\passumption{}{\pobject{a}{A}}
	}
}
& \hspace{1cm}
\infer[\pstop]{\pobject{
\ppair{a}{b}
}{
\pexists{a}{A}{B(a)}
}}
{
	\pobject{a}{A}
	&
	\pobject{b}{B(a)}
}
\end{matrix}
\]
One particular type that we shall meet often in this paper is 
$$\pforall{a}{A}{P(a) \pimplies \pexists{b}{B}{Q(a, b)}} \pcomma$$ which defines the specification of a transformation that takes a source class $A$ to a target class $B$, subject to precondition $P(a)$ and postcondition $Q(a, b)$. Note that types $A \land B$ and $A \pimplies B$ are special cases of types $\pforall{a}{A}{B(a)}$ and $\pexists{a}{A}{B(a)}$ where $B$ is independent of $a$. 
A term $\lambda a\colon A.b$ of type $A \rightarrow B$ represents a function from $A$ to $B$.

Application is written simply as juxtaposition: 
\[
\begin{matrix}
\infer[]{\pobject{(t\ s)}{B(s)}}
{
	\pobject{t}{\forall a\colon A.B(a)}
	&
	\pobject{s}{A}
}
\end{matrix}
\pstop
\]
Further, the reduction relation is generated by the $\beta$-rule:
$$(\lambda a\colon A. t)s \rightarrow t\{a \mapsto s\}$$
 The reflexive and transitive closure of the one-step reduction
relation $\rightarrow$ is denoted by $\pcomputesto$.

We shall also add to our language a predicate $=$, which we can use to
build dependent types like $x = 0$, where $\pobject{x}{\pnat}$.
When $0$ is substituted for $x$, the type becomes $0 = 0$, which is
inhabited by $r(0)$ (see~\cite{thompson-1999} for more details); when
$1$ is substituted for $x$, the type becomes $1 = 0$, which is uninhabited.
Lastly, we shall add the type $[E]$ of lists of elements
of type $E$ to our language, and two distinguished types $Set$ and $Prop$, which will be
used to classify types.

\section{Type Theory for Model Transformations}
\label{sec:3-ttfmt}

In this section, we formalise UML classes and objects using constructive type theory.

\begin{definition}
A UML class $C$ is encoded as a type, that is,  an inhabitant of type $\psetuni$; and a UML object of $C$ is encoded as an inhabitant of type $C$. Furthermore, a base attribute of $C$ is encoded as an inhabitant of type $C \pimplies \tau_1$, where $\tau_1$ is a ground type, e.g. $\pnat$; and a referential attribute of $C$ is encoded as an inhabitant of type $C \pimplies \tau_2$, where $\tau_2$ is the type of some other UML class or class list. 
\end{definition}

We shall assume that every UML class $C$ has a single base attribute $\pid{C}$ of type $\pnat$, and as many referential attributes as it needs to encode the relationships in which it participates. For example, if C is linked by a one-valued relationship to UML class D, and a many-valued relationship to UML class E, then the rule for constructing the inhabitants of $C$ is as follows, where $\pbuild{C}$ denotes an anonymous constructor of $C$.
\[
\infer[\pstop]{
\pobject{
\pappnpfour{\pbuild{C}}{n}{d}{l}
}{C}}
{
	\pobject{n}{\pnat}
	&
	\pobject{d}{D}
	&
	\pobject{l}{\plist{E}}
}
\]

The judgement $\pobject{a}{A}$ admits several different readings: $a$ is an inhabitant of type $A$ (as above), $a$ is a program whose specification is $A$ (which may be that of a model transformation), and $a$ is a proof of proposition $A$ (which may be that of a precondition). In the last reading, $A$ is defined to be an inhabitant of type $\ppropuni$, where $A$ is considered to be true if and only if it is inhabited. The relationship between propositions and types, which was first discovered by Curry \cite{curry-1958} and later extended by Howard \cite{howard-1980}, is known as the Curry-Howard isomorphism.

In this paper, we describe a technique to derive proofs of potentially large ordered model transformations. To illustrate the ideas underlying this technique, we consider first a simple model transformation, where each object of class $A$ (see Fig. \ref{fig:3-asmt.red}) 
is transformed into an object of class $P$, subject to a precondition $Pre_A$ of type $A \pimplies \ppropuni$ and a postcondition $Post_P$ of type $A \pimplies P \pimplies \ppropuni$.
\footnote{
Preconditions serve several purposes. First, to allow a choice of rules in different cases, e.g. by checking that a class is a root class, if root classes are transformed
by a different rule to non-root classes. Second, to ensure that a postcondition is well-defined, e.g. by insisting that $x \ge 0$ if the postcondition takes the square root of $x$. Third, to ensure that only certain source elements are transformed, e.g. by checking that a class is persistent, if only persistent classes are mapped to database tables. In the first and second cases, we might expect the precondition to contribute to the proof of the postcondition.
}
\begin{figure}
\centerline{\xymatrix{
A \ar@{|-->}[rrr]^<<<<{\ppre{A}}^>>>>{\ppost{P}} &&& P
}}
\caption{A transformation between $A$ and $P$.}
\label{fig:3-asmt.red}
\end{figure}
The specification of the transformation is formalised as a type, i.e.
\begin{equation}
\pforall{a}{A}{
\pappnp{\ppre{A}}{a} \pimplies \pexists{p}{P}{
\pappnpthree{\ppost{P}}{a}{p}
}
} \pcomma
\end{equation}
and its proof is given by
\predefaz
\def\pA{
\pobject{
\plambda{a}{
\plambda{h}{
\ppair{\pwitness{p}}{u}
}
}
}{
\pforall{a}{A}{
\pappnp{Pre}{a} \pimplies \pexists{p}{P}{
\pappnpthree{Post}{a}{p}
}
}
}
}
\def\pB{
\pobject{
\plambda{h}{
\ppair{\pwitness{p}}{u}
}
}{
\pappnp{Pre}{a} \pimplies \pexists{p}{P}{
\pappnpthree{Post}{a}{p}
}
}
}
\def\pC{
\pobject{
\ppair{\pwitness{p}}{u}
}{
\pexists{p}{P}{
\pappnpthree{Post}{a}{p}
}
}
}
\def\pD{
\pobject{u}{
\pappnpthree{Post}{a}{\pwitness{p}}
}
}
\def\pE{
\pobject{a}{A}
}
\def\pF{
\pobject{h}{\pappnp{Pre}{a}}
}
\begin{equation}
\label{equation.a}
\infer[\pforallirule{1} \pcomma]{\pA}
{
	\infer[\parrowirule{2}]{\pB}
	{
		\infer[\pexistsirule]{\pC}
		{
			\infer*[Hole]{\pD}
			{
				\passumption{1}{\pE}
				&
				\passumption{2}{\pF}
			}
		}
	}
}
\end{equation}
i.e. a function that takes an object $a$ of $A$ and a proof $h$ of $\pappnp{Pre_A}{a}$, and returns as a pair the corresponding object $\pwitness{p}$ of $P$ and a proof $u$ of $\pappnpthree{Post_P}{a}{\pwitness{p}}$.
There is a hole in the proof above  because the transformation is under specified. However, given suitable definitions of $A$, $P$, $\ppre{A}$ and $\ppost{P}$, the hole could be filled and the proof completed. Furthermore, given a second transformation with a different set of definitions of $A$, $P$, $\ppre{A}$ and $\ppost{P}$, we could apply the same procedure. However, the proofs would be so similar, at least in outline, that it should be possible to capture them all in a parametrised proof, by quantifying over all source and target classes, pre and postconditions, and proofs of holes, in the specification of the transformation. Based on this idea, we define the following correctness condition.

\begin{definition}[Correct Model Transformation] 
A correct model transformation from $X$ to $Y$
should ensure that for each $x$ in $X$ that satisfies the
precondition there is a $y$ in $Y$ that satisfies the postcondition. 
This is formalised using the following type:
\begin{gather}
\pforall{X}{\psetuni}{
\pforall{Y}{\psetuni}{
\pforall{\ppre{}}{X \pimplies \ppropuni}{
\pforall{\ppost{}}{X \pimplies Y \pimplies \ppropuni}{}
}
}
}\notag\\
\pforall{f}{X \pimplies Y}{
\pforall{Hole}{
(\pforall{x}{X}{\pappnp{\ppre{}}{x} \pimplies
\pappnpthree{\ppost{}}{x}{\papp{f}{x}}
})
}{}
}
\notag\\
\pforall{x}{X}{
\pappnp{Pre}{x} \pimplies \pexists{y}{Y}{
\pappnpthree{Post}{x}{y}
}
}
\label{equ:3-asmt.green}
\pstop
\end{gather}
\end{definition}

The proof of \eqref{equ:3-asmt.green} is little more than two eliminations and a sequence of introductions, i.e.
 \predefaz
 \def\pA{
 \pforall{X}{Set}{
 \pforall{Y}{Set}{\dots}
 }
 }
 \def\pB{
 \pforall{Y}{Set}{
 \pforall{Pre}{X \pimplies Prop}{\dots}
 }
 }
 \def\pC{
 \pforall{Pre}{X \pimplies Prop}{
 \pforall{Post}{X \pimplies Y \pimplies Prop}{\dots}
 }
 }
 \def\pD{
 \pforall{Post}{X \pimplies Y \pimplies Prop}{
 \pforall{f}{X \pimplies Y}{\dots}
 }
 }
 \def\pE{
 \pforall{f}{X \pimplies Y}{
 \pforall{Hole}{
 (\pforall{x}{X}{\pappnp{Pre}{x} \pimplies
 \pappnpthree{Post}{x}{\papp{f}{x}}
 })
 }{\dots}
 }
 }
 \def\pF{
 \pforall{Hole}{
 (\pforall{x}{X}{\pappnp{Pre}{x} \pimplies \pappnpthree{Post}{x}{\papp{f}{x}}})
 }{
 \pforall{x}{X}{\dots}
 }
 }
 \def\pG{
 \pforall{x}{X}{
 \pappnp{Pre}{x} \pimplies \pexists{y}{Y}{
 \pappnpthree{Post}{x}{y}
 }
 }
 }
 \def\pH{
 \pappnp{Pre}{x} \pimplies \pexists{y}{Y}{
 \pappnpthree{Post}{x}{y}
 }
 }
 \def\pI{
 \pexists{y}{Y}{
 \pappnpthree{Post}{x}{y}
 }
 }
 \def\pJ{
 \pappnpthree{Post}{x}{\papp{f}{x}}
 }
 \def\pK{
 \pappnp{Pre}{x} \pimplies \pJ
 }
 \def\pL{
 \pobject{h}{\pappnp{Pre}{x}}
 }
 \def\pM{
 \pobject{Hole}{
 (\pforall{x}{X}{\pappnp{Pre}{x} \pimplies
 \pappnpthree{Post}{x}{\papp{f}{x}}})
 }
 }
 \def\pN{
 \pobject{x}{X}
 }
 \def\pO{
 \pobject{X}{Set}
 }
 \def\pP{
 \pobject{Y}{Set}
 }
 \def\pQ{
 \pobject{Pre}{X \pimplies Prop}
 }
 \def\pR{
 \pobject{Post}{X \pimplies Y \pimplies Prop}
 }
 \def\pS{
 \pobject{f}{X \pimplies Y}
 }
 \begin{equation}
 \begin{small}
 \infer[\pforallirule{1} \pstop]{\pA}
 {
 	\infer[\pforallirule{2}]{\pB}
 	{
 		\infer[\pforallirule{3}]{\pC}
 		{
 			\infer[\pforallirule{4}]{\pD}
 			{
 				\infer[\pforallirule{5}]{\pE}
 				{
 					\infer[\pforallirule{6}]{\pF}
 					{
 						\infer[\pforallirule{7}]{\pG}
 						{
 							\infer[\parrowirule{8}]{\pH}
 							{
 								\infer[\pexistsirule]{\pI}
 								{
 									\infer[\parrowerule]{\pJ}
 									{
 										\infer[\pforallerule]{\pK}
 										{
 											\infer*[]{\passumption{6}{\pM}}
 											{
 												\begin{matrix}
 													\passumption{1}{\pO}\\
 													\passumption{2}{\pP}\\
 													\passumption{3}{\pQ}\\
 													\passumption{4}{\pR}\\
 													\passumption{5}{\pS}
 												\end{matrix}
 											}
 											&
 											\passumption{7}{\pN}
 										}
 										&
 										\passumption{8}{\pL}
 									}
 								}
 							}
 						}
 					}
 				}			
 			}
 		}
 	}
 }
 \end{small}
 \end{equation}

The fixed outline shape of the proof is captured by rules $\pexistsirule$ to $\pforallirule{7}$, and the variable proof of the hole is captured by assumption 6. Furthermore, the function $K$ defined below can easily be shown to inhabit \eqref{equ:3-asmt.green}.
\begin{equation}
\label{equ:3-asmt.blue}
K \df \plambda{X}{
\plambda{Y}{
\plambda{\ppre{}}{
\plambda{\ppost{}}{
\plambda{f}{
\plambda{Hole}{
\plambda{x}{
\plambda{h}{
\ppair{\papp{f}{x}}{u}
}
}
}
}
}
}
}
} \pstop
\end{equation}
Note that 
the arguments $X$ and $Y$ are arbitrary source and target classes; $\ppre{}$ and $\ppost{}$
are arbitrary pre and postconditions; $f$ is a function that maps source objects to target objects; $Hole$ is a proof of the hole (see \eqref{equation.a}); $x$ is a source object; and $h$ is a proof that the precondition holds on the source object. $K$ returns a target object $\papp{f}{x}$, and a proof $u$ that the postcondition holds on the source and target objects.

We shall now apply $K$ to a particular transformation, i.e. the one between $A$ and $P$. Let $\ppre{A}$ be a predicate that holds on all objects of $A$, and let $\ppost{P}$ be a predicate that holds on all objects of $A$ and $P$ which have the same base attribute values. Formally, let
\begin{align*}
\ppre{A} &\df \plambda{a}{\top}\\
\ppost{P} &\df \plambda{a}{
\plambda{p}{
(\pappnp{\pbaseat{A}}{a} = \pappnp{\pbaseat{P}}{p})
}
} \pstop
\end{align*}
Now, if
\[
f_A \df \plambda{a}{
\pappnp{\pbuild{P}}{
\papp{\pid{A}}{a}
}
} \pcomma
\]
then the proof of the hole is
\predefaz
\def\pA{
\pobject{
\plambda{a}{
\plambda{h}{
\preflex{\pappnp{\pid{A}}{a}}
}
}
}{
\pforall{a}{A}{
\pappnp{\ppre{A}}{a} \pimplies \pappnpthree{\ppost{P}}{a}{\papp{f_A}{a}}
}
}
}
\def\pB{
\pobject{
\plambda{a}{
\plambda{h}{
\preflex{\pappnp{\pid{A}}{a}}
}
}
}{
\pforall{a}{A}{
\top \pimplies \pappnp{\pid{A}}{a} = \pappnp{\pid{A}}{a}
}
}
}
\def\pC{
\pobject{
\plambda{h}{\preflex{\pappnp{\pid{A}}{a}}}
}{
\top \pimplies \pappnp{\pid{A}}{a} = \pappnp{\pid{A}}{a}
}
}
\def\pD{
\pobject{\preflex{\pappnp{\pid{A}}{a}}}{
\pappnp{\pid{A}}{a} = \pappnp{\pid{A}}{a}
}
}
\def\pE{
\pobject{\pappnp{\pid{A}}{a}}{\pnat}
}
\def\pF{
\pobject{\pid{A}}{
A \pimplies \pnat
}
}
\def\pG{
\pobject{a}{A}
}
\def\pH{
\pobject{h}{\top}
}
\begin{equation*}
\infer[\pdefrule \pstop]{\pA}
{
	\infer[\pforallirule{1}]{\pB}
	{
		\infer[\parrowirule{2}]{\pC}
		{
			\infer[\piirule]{\pD}
			{
				\infer[\parrowerule]{\pE}
				{
					\penviron{\pF}
					&
					\infer*[]{\passumption{1}{\pG}}
					{
						\passumption{1}{\pG}
						&
						\passumption{2}{\pH}
					}
				}
			}
		}
	}
}
\end{equation*}
Furthermore, if
\[
Hole_A \df \plambda{a}{
\plambda{h}{
\preflex{\pappnp{\pid{A}}{a}}
}
} \pcomma
\]
then
\[
\pappnpfour{K}{A}{P}{
\pappnpfour{\ppre{A}}{\ppost{P}}{f_A}{
\pappnpthree{Hole_A}{\papp{\pbuild{A}}{1}}{Triv}
}
} \pcomputesto \ppair{
\papp{\pbuild{P}}{1}
}{
\preflex{\pappnp{\pid{A}}{
\papp{\pbuild{A}}{1}
}}
} \pstop
\]
Therefore, $\pappnp{\pbuild{P}}{1}$ is the transform of $\pappnp{\pbuild{A}}{1}$, and $\preflex{\pappnp{\pid{A}}{\papp{\pbuild{A}}{1}}}$ is the term that proves it.

In the next sections we generalise these ideas to families of ordered model transformations.

\section{Ordered Model Transformations}
\label{sec:4-omt}

In the previous section, we showed how the specification of a small transformation could be abstracted into the specification of a family of transformations, by quantifying over all possible source and target classes, pre and postconditions and holes. The proof that resulted contained a fixed part, common to all members of the family, and a variable part, specific to a particular member. In this section, we extend these ideas to ordered model transformations in general.


There are two kinds of ordered transformations: totally and partially ordered.  Informally, a totally ordered set of transformations is one that has the shape of a ladder, in which the source and target models are the verticals, and the transformations are the rungs. A partially ordered set of transformations has the shape of a tree of ladders, in which the branches between nodes are totally ordered transformations. We give examples before presenting the formal definitions.

\begin{figure}
\centerline{\xymatrix{
A
\ar@{|-->}[rrr]^<<<<{\ppre{A}}^>>>>{\ppost{P}}
\ar@{->}[d]^{R_1} &&& 
P
\ar@{->}[d]^{S_1}\\
B
\ar@{|-->}[rrr]^<<<<{\ppre{B}}^>>>>{\ppost{Q}} &&& Q
}}
\caption{A two-runged transformation}
\label{fig:4-omt.red}
\end{figure}
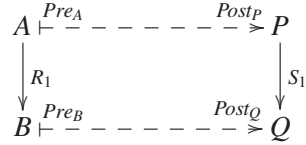

Consider the two-runged transformation in Fig. \ref{fig:4-omt.red}, in which $A$ is transformed to $P$ subject to conditions $\ppre{A}$ and $\ppost{P}$, and $B$ is transformed to $Q$ subject to conditions $\ppre{B}$ and $\ppost{Q}$. If an object $a$ of $A$ is transformed to an object $p$ of $P$, then the object $\pappnp{R_1}{a}$ of $B$ is transformed to the object $\pappnp{S_1}{p}$ of $Q$. In other words, the transformation of $B$ to $Q$ is nested within the transformation of $A$ to $P$. As before, we shall define the specification of the transformation as a type, outline its proof, and abstract over classes, conditions and holes. The transformation along each rung is similar to the one between $A$ and $P$ earlier. However, the transformation between the verticals (relationships) is new.

The specification of the transformation that is depicted in Figure~\ref{fig:4-omt.red}
 is formalised as a type as
follows (for ease of readability, we write one rung per line).
\begin{gather}
\pforall{a}{A}{
\pappnp{\ppre{A}}{a} \pimplies \pexists{p}{P}{(
\pappnpthree{\ppost{P}}{a}{p}
}
} \pland \notag\\
\pforall{b}{B}{
\pappnp{\ppre{B}}{b} \pland (\pappnp{R_1}{a}) = b \pimplies
\pexists{q}{Q}{
\pappnpthree{\ppost{Q}}{b}{q} \pland (\pappnp{S_1}{p}) = q
}
}) \pstop
\label{equ:4-omt.orange}
\end{gather}
The transformation along the second rung is nested within the first rung, and has stronger pre and postconditions in virtue of the connectivity constraints placed on objects of $B$ and $Q$. Note that a third or fourth rung would have the same shape as the second rung.

In outline, the proof of \eqref{equ:4-omt.orange} is
\predefaz
\def\pA{
\pforall{a}{A}{
\pappnp{\ppre{A}}{a} \pimplies \pexists{p}{P}{
\pappnpthree{\ppost{P}}{a}{p}
} \pland \dots
}}
\def\pB{
\psubstitute{(
\pappnpthree{\ppost{P}}{a}{p} \pland
\pforall{b}{B}{
\pappnp{\ppre{B}}{b} \pland \pappnp{R_1}{a} = b \pimplies
\dots}
)}{\pwitness{p}}{p}
}
\def\pC{
\pappnpthree{\ppost{P}}{a}{\pwitness{p}}
}
\def\pD{
\pobject{a}{A}
}
\def\pE{
\pappnp{\ppre{A}}{a}
}
\def\pF{
\psubstitute{(
\pforall{b}{B}{
\pappnp{\ppre{B}}{b} \pland \pappnp{R_1}{a} = b \pimplies
\dots}
)}{\pwitness{p}}{p}
}
\def\pG{
\psubstitute{(
\pappnpthree{\ppost{Q}}{b}{q} \pland \pappnp{S_1}{\pwitness{p}} = q
)}{\pwitness{q}}{q}
}
\def\pH{
\pappnpthree{\ppost{Q}}{b}{\pwitness{q}}
}
\def\pI{
\pappnp{\ppre{B}}{b}
}
\def\pJ{
\pobject{b}{B}
}
\def\pK{
\pappnp{\ppre{B}}{b} \pland \pappnp{R_1}{a} = b
}
\def\pL{
\pappnp{S_1}{\pwitness{p}} = \pwitness{q}
}
\def\pM{
\pappnp{R_1}{a} = b
}
\begin{equation*}
\infer*[Fixed]{\pA}
{
	\infer[\pandirule]{\pB}
	{
		\infer*[Hole_A]{\pC}
		{
			\begin{matrix}
				\passumption{}{\pD}\\
				\passumption{}{\pE}
			\end{matrix}
		}
		&
		\infer*[Fixed]{\pF}
		{
			\infer[\pandirule]{\pG}
			{
				\infer*[Hole_B]{\pH}
				{
					\infer[\panderule{1}]{\penviron{\pI}}
					{
						\begin{matrix}
							\passumption{}{\pD}\\
							\passumption{}{\pJ}\\
							\passumption{}{\pK}
						\end{matrix}
					}
				}
				&
				\infer*[Com_A]{\pL}
				{
					\infer[\panderule{2}]{\penviron{\pM}}
					{
						\begin{matrix}
							\passumption{}{\pD}\\
							\passumption{}{\pJ}\\
							\passumption{}{\pK}
						\end{matrix}
					}
				}
			}
		}
	}
}
\end{equation*}
The fixed parts of the proof exist in virtue of the structure of the specification, whereas the variable parts exist in virtue of its under-specification. The variable parts are either of the $Hole$ kind, i.e. proofs that postconditions are derived from preconditions, or the $Com$ kind, i.e. proofs that adjacent rungs are linked.

Quantifying over all variables in \eqref{equ:4-omt.orange}, including proofs of variable parts, and changing the names of bound variables where appropriate, we obtain the specification of an arbitrary two-runged transformation.

\begin{definition}[Two-Runged Model Transformation]
An arbitrary two-runged model transformation is formalised in constructive type theory by the following type:
\begin{gather}
\pforall{X}{\psetuni}{
\pforall{Y}{\psetuni}{
\pforall{\ppre{}}{X \pimplies \ppropuni}{
\pforall{\ppost{}}{X \pimplies Y \pimplies \ppropuni}{}
}
}
}\notag\\
\pforall{f}{X \pimplies Y}{
\pforall{Hole}{
(\pforall{x}{X}{\pappnp{\ppre{}}{x} \pimplies
\pappnpthree{\ppost{}}{x}{\papp{f}{x}}
})
}{}
}\notag\\
\pforall{X'}{\psetuni}{
\pforall{Y'}{\psetuni}{
\pforall{\ppre{}'}{X' \pimplies \ppropuni}{
\pforall{\ppost{}'}{X' \pimplies Y' \pimplies \ppropuni}{}
}
}
}\notag\\
\pforall{f'}{X' \pimplies Y'}{
\pforall{Hole'}{
(\pforall{x'}{X'}{\pappnp{\ppre{}}{x'} \pimplies
\pappnpthree{\ppost{}}{x'}{\papp{f'}{x'}}
})
}{}
}\notag\\
\pforall{R}{X \pimplies X'}{
\pforall{S}{Y \pimplies Y'}{}
}\notag\\
\pforall{Com}{(
\pforall{x}{X}{
\pappnp{S}{\papp{f}{x}} = \pappnp{f'}{\papp{R}{x}}
}
)}{}
\notag\\
\pforall{x}{X}{
\pappnp{\ppre{}}{x} \pimplies \pexists{y}{Y}{
\pappnpthree{\ppost{}}{x}{y}
}
} \pland \notag\\
\pforall{x'}{X'}{
\pappnp{\ppre{}'}{x'} \pland \pappnp{R}{x} = x' \pimplies
\pexists{y'}{Y'}{
\pappnpthree{\ppost{}'}{x'}{y'} \pland \pappnp{S}{y} = y'
}
} \pstop
\label{equ:4-omt.yellow}
\end{gather}
\end{definition}
This formalisation is similar to the one in \eqref{equ:3-asmt.green} except that it also quantifies over $Com$, i.e. a proof that starting from every object $x$ of $X$ ($X$ being the source end of the first rung) and navigating to some object $y'$ of $Y'$ ($Y'$ being the target end of the second rung), first via $f$ and $S$, and then via $R$ and $f'$, the same $y'$ is obtained.
\begin{figure}
\centerline{\xymatrix{
x
\ar@{-->}[rr]^{f}
\ar@{->}[d]^{R}
\ar@{}[drr]|{Com} && 
\pappnp{f}{x}
\ar@{->}[d]^{S}\\
\pappnp{R}{x}
\ar@{-->}[rr]^{f'} && 
\pappnp{S}{\papp{f}{x}} = \pappnp{f'}{\papp{R}{x}}
}}
\caption{The commutative square $Com$.}
\label{fig:4-omt.violet}
\end{figure}
The commutative square, after which $Com$ is named, is shown in Fig.~\ref{fig:4-omt.violet}. Furthermore, given a proof of $Com$, it is a trivial matter to prove that the rungs are linked, i.e.
\predefaz
\def\pA{
\pappnp{S}{\papp{f}{x}} = \pappnp{f'}{x'}
}
\def\pB{
\pappnp{S}{\papp{f}{x}} = \pappnp{f'}{\papp{R}{x}}
}
\def\pC{
\pobject{x}{X}
}
\def\pD{
\pobject{Com}{
\pforall{x}{X}{
\pappnp{S}{\papp{f}{x}} = \pappnp{f'}{\papp{R}{x}}
}
}
}
\def\pE{
\pappnp{R}{x} = x'
}
\begin{equation*}
\infer[\pierule \pstop]{\pA}
{
	\infer[\pforallerule]{\pB}
	{
		\passumption{}{\pC}
		&
		\passumption{}{\pD}
	}
	&
	\penviron{\pE}
}
\end{equation*}

\begin{property}
The function that inhabits \eqref{equ:4-omt.yellow} is
$$\lambda X\: Y \: Pre \: Post\: f\: Hole\: X'\: Y'\: Pre'\: Post'\: f'\: Hole'\: R\: S\: Com\: x\: h.\ppair{\papp{f}{x}}{u}$$
\end{property}

\begin{proof}
Direct, using the typing rules given in Section~\ref{sec:2-p}.
Note that the first element of the output pair is the result of applying the root function $f$ to an arbitrary root object $x$, reflecting the fact that an ordered transformation is essentially a transformation between root classes.
\end{proof}

We are now ready to formalise the notion of ordered model transformation. 
First we consider totally  ordered transformations, and later we extend the results to partially ordered transformations.

\subsection{Totally Ordered Transformations}

In order to formalise totally ordered model transformations, we will define a dependent type
$\pappnpfour{T}{X}{Y}{f}$, where $T$ is the type name, and $X$, $Y$ and $f$ are the parameters on which it depends, i.e. root source class, root target class and root function respectively. A totally ordered transformation is an inhabitant of this dependent type.   The inhabitants of $T$ are defined inductively, in much the same way as $\pnat$, by means of a base rule and a step rule. 

\begin{definition}(Type $\pappnpfour{T}{X}{Y}{f}$)
\label{def:TXYf}
 The type $T$ is defined inductively, with a rule defining the base case and a rule defining the inductive step, as follows:

\predefaz
\def\pA{
\ar@{.>}[rr]^{f}
\ar@{->}[d]^{R}
\ar @{} [drr] |{Com}
}
\def\pB{
\ar@{->}[d]^{S}
}
\def\pC{
\ar@{.>}[rr]^{f}
\ar@{->}[d]^{R}
\ar @{} [drr] |{Com}
}
\def\pD{
\ar@{->}[d]^{S}
}
\def\pE{
\ar@{-->}[rr]^{f'}_{Hole'}^<<<{Pre'}^>>>{Post'}
\ar@{.}[d]
\ar @{} [drr] |{
\pobject{t'}{
\pappnp{T}{\pappnp{X'}{\pappnp{Y'}{f'}}}}
}
}
\def\pF{
\ar@{.}[d]
}
\def\pG{
\ar@{-->}[rr]^{f'}_{Hole'}^<<<{Pre'}^>>>{Post'}
}
\def\pH{
}
\def\pI{
\ar@{.}[rr]
}
\def\pJ{
}

\predefaz
\def\pA{
\pobject{
\pappnp{T_{Base}}{\pappnp{X}{\pappnp{Y}{\pappnp{f}{\pappnp{X'}{\pappnp{Y'}{\pappnp{f'}{\pappnp{Pre'}{\pappnp{Post'}{\pappnp{Hole'}{\pappnp{R}{\pappnp{S}{Com}}}}}}}}}}}}
}{\pappnp{T}{\pappnp{X}{\pappnp{Y}{f}}}}
}
\def\pB{
\begin{array}{l}
\pobject{X}{\psetuni}\quad \pobject{Y}{\psetuni}
\quad \pobject{f}{X \pimplies Y}\\
\pobject{X'}{\psetuni}\quad \pobject{Y'}{\psetuni}
\quad \pobject{f'}{X' \pimplies Y'}\\
\pobject{Pre'}{X' \pimplies \ppropuni}
\quad \pobject{Post'}{X' \pimplies Y' \pimplies \ppropuni}\\
\pobject{Hole'}{\pforall{x'}{X'}{
\pappnp{Pre'}{x'} \pimplies \pappnp{Post'}{\pappnp{x'}{
\papp{f'}{x'}
}}
}}\\
\pobject{R}{X \pimplies X'}
\quad \pobject{S}{Y \pimplies Y'}
\quad \pobject{Com}{\pforall{x}{X}{
\pappnp{f'}{\papp{R}{x}} = \pappnp{S}{\papp{f}{x}}
}}
\end{array}
}
\begin{equation*}
\label{equ:4-omt.indigo}
\infer[(TI_1)]{\pA}
{
	\pB
}
\end{equation*}

\predefaz
\def\pA{
\pobject{
\pappnp{T_{Step}}{\pappnp{X}{\pappnp{Y}{\pappnp{f}{\pappnp{X'}{\pappnp{Y'}{\pappnp{f'}{\pappnp{Pre'}{\pappnp{Post'}{\pappnp{Hole'}{\pappnp{R}{\pappnp{S}{\pappnp{Com}{t'}}}}}}}}}}}}}
}{\pappnp{T}{\pappnp{X}{\pappnp{Y}{f}}}}
}
\def\pB{
\begin{array}{l}
\pobject{X}{\psetuni}\quad \pobject{Y}{\psetuni}
\quad \pobject{f}{X \pimplies Y}\\
\pobject{X'}{\psetuni}\quad \pobject{Y'}{\psetuni}
\quad \pobject{f'}{X' \pimplies Y'}\\
\pobject{Pre'}{X' \pimplies Prop}
\quad \pobject{Post'}{X' \pimplies Y' \pimplies Prop}\\
\pobject{Hole'}{\pforall{x'}{X'}{
\pappnp{Pre'}{x'} \pimplies \pappnp{Post'}{\pappnp{x'}{
\papp{f'}{x'}
}}
}}\\
\pobject{R}{X \pimplies X'}
\quad \pobject{S}{Y \pimplies Y'}
\quad \pobject{Com}{\pforall{x}{X}{
\pappnp{f'}{\papp{R}{x}} = \pappnp{S}{\papp{f}{x}}
}}\\
\pobject{t'}{
\pappnp{T}{\pappnp{X'}{\pappnp{Y'}{f'}}}
}
\end{array}
}
\begin{equation*}
\label{equ:4-omt.violet}
\infer[(TI_2)]{\pA}
{
	\pB
}
\end{equation*}

\end{definition}

The base rule constructs a transformation of the kind shown in Fig. \ref{fig:4-omt.orange} (left). Note that the root hole and root pre and postconditions (to clarify, the root is at the top) are not part of the construction. The step rule
constructs the successor of a transformation $t'$, of the kind shown in Fig. \ref{fig:4-omt.orange} (middle). Again, the root hole and root pre and postconditions are not part of the construction.
\predefaz
\def\pA{
\ar@{.>}[rr]^{f}
\ar@{->}[d]^{R}
\ar @{} [drr] |{Com}
}
\def\pB{
\ar@{->}[d]^{S}
}
\def\pC{
\ar@{.>}[rr]^{f}
\ar@{->}[d]^{R}
\ar @{} [drr] |{Com}
}
\def\pD{
\ar@{->}[d]^{S}
}
\def\pE{
\ar@{-->}[rr]^{f'}_{Hole'}^<<<{Pre'}^>>>{Post'}
\ar@{.}[d]
\ar @{} [drr] |{
\pobject{t'}{
\pappnp{T}{\pappnp{X'}{\pappnp{Y'}{f'}}}}
}
}
\def\pF{
\ar@{.}[d]
}
\def\pG{
\ar@{-->}[rr]^{f'}_{Hole'}^<<<{Pre'}^>>>{Post'}
}
\def\pH{
}
\def\pI{
\ar@{.}[rr]
}
\def\pJ{
}
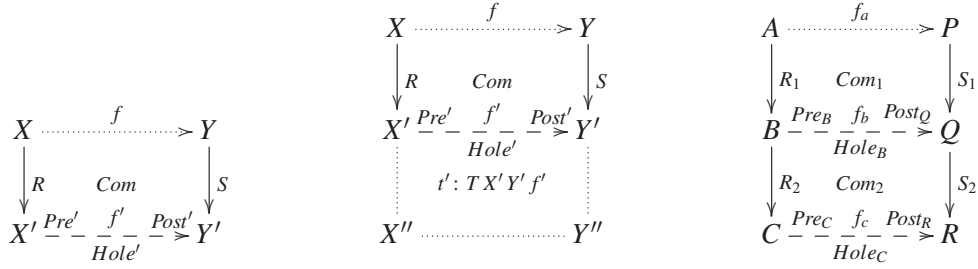
\begin{figure}
\centerline{\xymatrix{
&& && 
X\pA && Y\pB &&
A
\ar@{.>}[rr]^{f_a}
\ar@{->}[d]^{R_1}
\ar @{} [drr] |{Com_1} && 
P
\ar@{->}[d]^{S_1}\\
X\pC && Y\pD && X'\pE && Y'\pF &&
B
\ar@{-->}[rr]^<<<<{\ppre{B}}^>>>>{\ppost{Q}}^{f_b}_{Hole_B}
\ar@{->}[d]^{R_2}
\ar @{} [drr] |{Com_2} && Q
\ar@{->}[d]^{S_2}
\\
X'\pG && Y'\pH && X''\pI && Y''\pJ &&
C
\ar@{-->}[rr]^<<<<{\ppre{C}}^>>>>{\ppost{R}}^{f_c}_{Hole_C} && R
}}
\caption{From left to right: representations of the base and step rules of type $T$, followed by an inhabitant of type $\pappnpfour{T}{A}{P}{f_a}$, namely $t_{AP}$.}
\label{fig:4-omt.orange}
\end{figure}

To construct an inhabitant of $T$, we first apply the base rule and then repeatedly apply the step rule. For example, the transformation $t_{AP}$ in Fig. \ref{fig:4-omt.orange} (right) is constructed as follows:
\begin{align*}
t_{BQ} &\df
\pappnp{T_{Base}}{
\pappnp{B}{
\pappnp{Q}{
\pappnp{f_b}{
\pappnp{C}{
\pappnp{R}{
\pappnp{f_c}{
\pappnp{Pre_C}{
\pappnp{Post_R}{
\pappnp{Hole_C}{
\pappnp{R_2}{
\pappnp{S_2}{
Com_2
}}}}}}}}}}}} \pcomma\\
t_{AP} &\df
\pappnp{T_{Step}}{
\pappnp{A}{
\pappnp{P}{
\pappnp{f_a}{
\pappnp{B}{
\pappnp{Q}{
\pappnp{f_b}{
\pappnp{Pre_B}{
\pappnp{Post_Q}{
\pappnp{Hole_B}{
\pappnp{R_1}{
\pappnp{S_1}{
\pappnp{Com_1}{
t_{BQ}
}}}}}}}}}}}}} \pstop
\end{align*}

By extension of \eqref{equ:4-omt.orange}, it would be easy to write down the specification of $t_{AP}$, including the root transformation we have so assiduously excluded. However, much more useful would be to write down a function that could compute it, not only for $t_{AP}$ but also for every other inhabitant of $\pappnpfour{T}{X}{Y}{f}$ as well. Such a function is given below.


\begin{definition}(Spec)
\label{def:spec}
\predefaz
\def\pA{
\pobject{Spec}{
\pforall{X}{\psetuni}{
\pforall{Y}{\psetuni}{
\pforall{f}{X \pimplies Y}{
\pappnp{T}{
\pappnp{X}{
\pappnp{Y}{f} \pimplies (X \pimplies Y \pimplies \ppropuni)
}
}
}
}
}
}
}
\def\pB{
\pappnp{Spec}{\pappnp{X}{\pappnp{Y}{\pappnp{f}{(
\pappnp{T_{Base}}{\pappnp{X}{\pappnp{Y}{\pappnp{f}{\pappnp{X'}{\pappnp{Y'}{\pappnp{f'}{\pappnp{Pre'}{\pappnp{Post'}{\pappnp{Hole'}{\pappnp{R}{\pappnp{S}{Com}}}}}}}}}}}}
)}}}}
}
\def\pC{
\plambda{\pobject{x}{X}}{\plambda{\pobject{y}{Y}}{
\pforall{x'}{X'}{
\pappnp{Pre'}{x'} \land x' = \pappnp{R}{x} \pimplies \pexists{y'}{Y'}{
\pappnp{Post'}{\pappnp{x'}{y'}} \land y' = \pappnp{S}{y}
}
}
}}
}
\def\pD{
\pappnp{Spec}{\pappnp{X}{\pappnp{Y}{\pappnp{f}{(
\pappnp{T_{Step}}{\pappnp{X}{\pappnp{Y}{\pappnp{f}{\pappnp{X'}{\pappnp{Y'}{\pappnp{f'}{\pappnp{Pre'}{\pappnp{Post'}{\pappnp{Hole'}{\pappnp{R}{\pappnp{S}{\pappnp{Com}{t'}}}}}}}}}}}}}
)}}}}
}
\def\pE{
\pappnp{Spec}{\pappnp{X'}{\pappnp{Y'}{\pappnp{f'}{\pappnp{t'}{\pappnp{x'}{y'}}}}}}
}
\begin{gather}
\pA\notag\\
\notag\\
\pB \df\notag\\
\pC\notag\\
\notag\\
\pD \df\notag\\
\pC \pland\notag\\
\pE
\label{equ:4-omt.black}
\pstop
\end{gather}
\end{definition}


Now, the specification of a transformation is an inhabitant of $\ppropuni$. However, $Spec$ returns an inhabitant of type $X \pimplies Y \pimplies \ppropuni$. Why? To allow it to be integrated with the root objects of type $X$ and $Y$, passed down to it by the root transformation.

In its general form, an ordered model transformation is formalised as follows.

\begin{definition}(Ordered Model Transformation)
\label{def:omt}
An ordered model transformation is an inhabitant of type
\def\pA{
\pforall{X}{Set}{
\pforall{Y}{Set}{
\pforall{f}{X \pimplies Y}{
\pforall{t}{
\pappnpfour{T}{X}{Y}{f}
}{
}
}
}
}
}
\def\pB{
\pforall{Pre}{X \pimplies Prop}{
\pforall{Post}{X \pimplies Y \pimplies Prop}{
}
}
}
\def\pC{
\pforall{Hole}{
\pforall{x}{X}{
\pappnp{Pre}{x} \pimplies \pappnp{Post}{\pappnp{x}{
\papp{f}{x}
}
}
}
}{}
}
\def\pD{
\pforall{x}{X}{
\pappnp{Pre}{x} \pimplies \pexists{y}{Y}{
\pappnp{Post}{\pappnp{x}{y}}
}
}
}
\def\pE{
\pappnp{Spec}{\pappnp{X}{\pappnp{Y}{\pappnp{f}{\pappnp{t}{\pappnp{x}{y}
}}}}}
}
\begin{gather}
\pA \notag\\
\pB \notag\\
\pC \notag\\
\pD \pland \pE
\label{equ:4-omt.green}
\pstop
\end{gather} 
\end{definition}

A proof of \eqref{equ:4-omt.green} (an inhabitant of the type) has the form:
\begin{equation}
\label{equ:4-omt.blue}
\plambda{X}{
\plambda{Y}{
\plambda{f}{
\plambda{t}{
\plambda{Pre}{
\plambda{Post}{
\plambda{Hole}{
\plambda{x}{
\plambda{h}{
\ppair{\papp{f}{x}}{u}
}
}
}
}
}
}
}
}
} \pcomma
\end{equation}
where $u$ is a proof of
\begin{equation*}
\pappnp{Post}{\pappnp{x}{\papp{f}{x}}} \pland
\pappnpseven{Spec}{X}{Y}{f}{t}{x}{\papp{f}{x}} \pstop
\end{equation*}

According to \eqref{equ:4-omt.blue}, if we could construct an inhabitant of type $T$ from suitable values of $Hole$ (a proof of the root hole), $X$ and $Y$ (the root classes), $f$ (the root function), and $Pre$ and $Post$ (the root conditions), then we could justifiable claim that $\ppair{\papp{f}{x}}{u}$ is a certified implementation of the transformation, for an arbitrary source object $x$. In other words, constructing a suitable value of $t$ is tantamount to proving the specification.

\subsection{Partially Ordered Transformations}

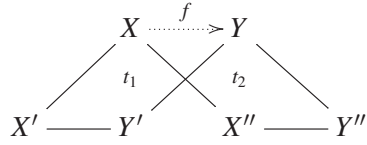
\begin{figure}
\centerline{\xymatrix{
& 
X \ar@{-}[ld] \ar@{-}[rd] \ar@{.>}[r]^{f} \ar@{}[d] | {t_1}
& 
Y \ar@{-}[ld] \ar@{-}[rd] \ar@{}[d] | {t_2}
& &\\
X' \ar@{-}[r]
& Y' &
X'' \ar@{-}[r]
& Y''
}}
\caption{A representation of the join rule of type $T$.}
\label{fig:4-omt.green}
\end{figure}

\begin{figure}[t]
\centerline{\xymatrix{
&& A \ar@{-}[d] \ar@{-->}[r]^{f_a} \ar@{}[dr]|{t_7}
 & P \ar@{-}[d]
 &&&\\
&& B \ar@{-}[dl] \ar@{-}[dr] \ar@{-->}[r]  \ar@{}[ddl]|{t_5} \ar@{}[drr]|{t_6}
& Q \ar@{-}[ddll] \ar@{-}[dr]
&&&\\
& C \ar@{-}[dl]
&& E \ar@{-}[d] \ar@{-->}[r] \ar@{}[dr]|{t_4}
& S \ar@{-}[d]
&&\\
D \ar@{-->}[r]
& R 
&& F \ar@{-}[dll] \ar@{-}[d] \ar@{-}[drr] \ar@{-->}[r] \ar@{}[dl]|{t_1} \ar@{}[dr]|{t_2} \ar@{}[drrr]|{t_3}
& T \ar@{-}[dll] \ar@{-}[d] \ar@{-}[drr]
&&\\
& G \ar@{-->}[r]
& U 
& H \ar@{-}[d] \ar@{-->}[r]
& V 
& J \ar@{-->}[r]
& W\\
&&& I 
&&&
}}
\caption{An example of a partially ordered transformation, in which $A, B, \dots, J$ are the source classes, $P, Q, \dots, W$ are the target classes, and $A$ is transformed to $P$, $B$ to $Q$ and so on. The transformation (minus the root artefacts) is given by $t_7$, where 
$t_1 \df \pappnp{T_{Base}}{\dots},\
t_2 \df \pappnp{T_{Base}}{\dots},\
t_3 \df \pappnp{T_{Base}}{\dots},\
t_{123} \df \pappnpthree{T_{Join}}{(
\pappnpthree{T_{Join}}{t_1}{t_2}
)}{t_3},\
t_4 \df \pappnpthree{T_{Step}}{\dots}{t_{123}},\
t_5 \df \pappnp{T_{Base}}{\dots},\
t_6 \df \pappnpthree{T_{Step}}{\dots}{t_4},\
t_{56} \df \pappnpthree{T_{Join}}{t_5}{t_6},$
and $t_7 \df \pappnpthree{T_{Step}}{\dots}{t_{56}}.$
The specification of the transformation is given by $\pappnpfive{Spec}{A}{P}{f_a}{t_7}$.
}
\label{fig:4-omt.blue}
\end{figure}
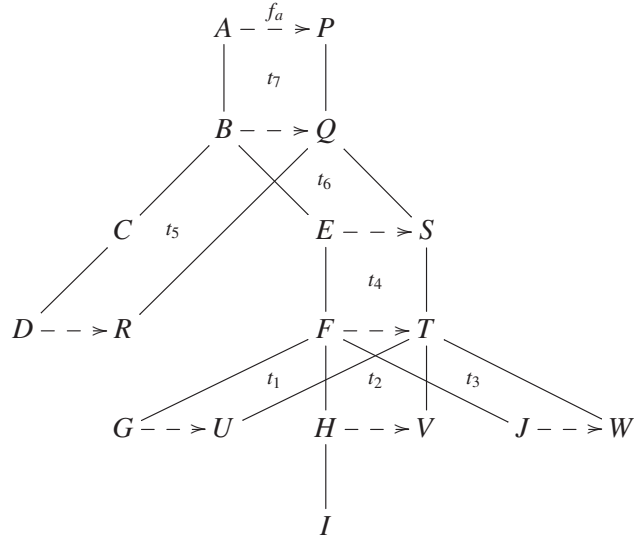

We generalise the construction to take into account the case where
transformations are partially ordered. Without loss of generality, we
assume that a partially ordered transformation can be constructed from
two ordered transformations, as shown in Fig. \ref{fig:4-omt.green}.
Thus, we extend the definition of $\pappnpfour{T}{X}{Y}{f}$ in Definition~\ref{def:TXYf}
 with a \textit{join} rule, i.e.
\predefaz
\def\pA{
\pobject{
\pappnpthree{T_{Join}}{t_1}{t_2}
}{
\pappnpfour{T}{X}{Y}{f}
}
}
\def\pB{
\pobject{t_1}{
\pappnpfour{T}{X}{Y}{f}
}
}
\def\pC{
\pobject{t_2}{
\pappnpfour{T}{X}{Y}{f}
}
}
\begin{gather*}
\infer[(TI_3) \pcomma]{\pA}
{
	\pB
	&
	\pC
}
\end{gather*}
and extend the definition of $Spec$ (Definition~\ref{def:spec})
 with a case for $T_{Join}$, which returns the conjunction of the specifications of $t_1$ and $t_2$, i.e.
\predefaz
\def\pF{
\pappnpfive{Spec\;}{X}{Y}{f}{(
\pappnpthree{T_{Join}}{t_1}{t_2}
)}
}
\def\pG{
\plambda{\pobject{x}{X}}{
\plambda{\pobject{y}{Y}}{
\pappnpseven{Spec}{X}{Y}{f}{t_1}{x}{y} \pland \pappnpseven{Spec}{X}{Y}{f}{t_2}{x}{y}
}
}
}
\begin{equation*}
\pF \df \pG \pstop
\end{equation*}
See Fig. \ref{fig:4-omt.blue} for an example.



\section{Concrete Example}
\label{sec:A-ce}

\begin{figure}[ht]
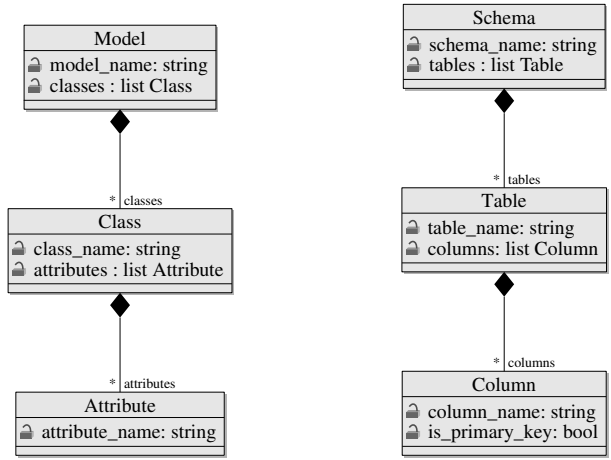

\begin{centering}
\includegraphics[scale=0.75]{UML-SQL-1.mps}
\hspace{2cm}
\includegraphics[scale=0.75]{UML-SQL-2.mps}
\caption{The UML and SQL models.}
\label{fig:A-ce.orange}
\end{centering}
\end{figure}

Consider a transformation between the UML and SQL models in Fig. \ref{fig:A-ce.orange}, in which
\begin{itemize}
\item
each model $m$ is mapped to a schema $s$ of the same name;
\item
each class $c$ in $m$ is mapped to a table $t$ in $s$ of the same name, and a primary key column in $t$ of the same name;
\item
each attribute in $c$ is mapped to a non-primary key column in $t$ of the same name;
\item
the mappings are unconditional, i.e. the preconditions always hold.
\end{itemize}

The specification of the transformation is given by
\begin{gather*}
\pforall{m}{Model}{\pappnp{Pre_{Model}}{m} \pimplies
\pexists{s}{Schema}{}}
\pappnpthree{Post_{Schema}}{m}{s} \pland\\
\pappnpseven{Spec}{Model}{Schema}{f_{Model}}{t_{Model-Schema}}{m}{s}
\pcomma
\end{gather*}
where
\[
\pobject{
t_{Model-Schema}
}{
\pappnpfour{T}{Model}{Schema}{f_{Model-Schema}}
}
\pstop
\]
If
\begin{align*}
m_1 &\df
\pappnpthree{\pbuild{Model}}{1}{
\plist{c_1, c_2, c_3}
}\\
c_1 &\df \pappnpthree{\pbuild{Class}}{2}{\plist{
\pappnp{\pbuild{Attribute}}{5},
\pappnp{\pbuild{Attribute}}{6},
\pappnp{\pbuild{Attribute}}{7}
}
}\\
c_1 &\df \pappnpthree{\pbuild{Class}}{3}{\plist{
\pappnp{\pbuild{Attribute}}{8}
}
}\\
c_3 &\df \pappnpthree{\pbuild{Class}}{4}{\pemptylist} \pcomma
\end{align*}
i.e. a model with 3 classes and 4 attributes, then
\[
\pappnpfour{K}{Model}{Schema}{
\pappnpfour{f_{Model}}{Pre_{Model}}{Post_{Schema}}{
\pappnpthree{Hole_{Model}}{m_1}{Triv}
}
}
\]
reduces to $\ppair{s_1}{p}$, where $s_1$ is given by
\begin{align*}
s_1 &\df
\pappnpthree{\pbuild{Schema}}{1}{
\plist{t_1, t_2, t_3}
}\\
t_1 &\df \pappnpthree{\pbuild{Table}}{2}{\plist{
\pappnpthree{\pbuild{Column}}{2}{true},
\pappnpthree{\pbuild{Column}}{5}{false},
\pappnpthree{\pbuild{Column}}{6}{false},
\pappnpthree{\pbuild{Column}}{7}{false},
}
}\\
t_1 &\df \pappnpthree{\pbuild{Table}}{3}{\plist{
\pappnpthree{\pbuild{Column}}{3}{true},
\pappnpthree{\pbuild{Column}}{8}{false}
}
}\\
t_3 &\df \pappnpthree{\pbuild{Table}}{4}{\plist{
\pappnpthree{\pbuild{Column}}{4}{true}
}} \pcomma
\end{align*}
i.e. a schema with 3 tables and 7 columns (3 of which are primary keys), and $p$ is an unspecified proof (through lack of space) that $s_1$ is indeed the transform of $m_1$.

\section{Related Work and Conclusions}
\label{sec:5-c}

A number of authors have attempted to provide a formal understanding of metamodelling and model transformations. Ruscio et al. have made some progress towards formalizing the KM3 metamodelling language using Abstract State Machines~\cite{ruscio-2006}; and Rivera and Vallecillo have exploited the class-based nature of the Maude specification language to formalize metamodels written in KM3 ~\cite{rivera-2007}. Further, a related algebraic approach is given by Boronat and Meseguer in~\cite{boronat-2008}. More recently, Calegari et al. \cite{Tasistro-Szasz-etc} proposed a framework for encoding models and metamodels in the Calculus of Inductive Constructions (CIC)~\cite{coq-2010, bertot-2004}, and in doing so showed how parts of the ATL model transformation language~\cite{jouault-2006} could be expressed in the CIC, including matched rules, helpers and expressions based on the Object Constraint Language (OCL) \cite{omg-ocl-2012}.

In this paper, we have shown how to assemble the proof of a potentially large ordered model transformation, by decomposing it into a number of smaller proofs which are easier to derive. 
\footnote{
The reader should note that this approach could never be fully automatised in virtue of the unlimited scope of pre and postconditions. However, once the smaller proofs are available, the process of assembling them into a proof of the whole could indeed be automatised. 
}
We focused on a particular kind of transformation with uniform characteristics, which we hope to extend to other kinds of transformations in the future, although we have already incorporated a number of additional variants into the scheme outlined above, including support for many-valued relationships, unmapped source classes (of which $C$ is an example in Fig. \ref{fig:4-omt.blue}) and multiple target classes.

In future work, we will extend the techniques to a larger class of model transformations, by abstracting over the non-hierarchical parts of models too. One way of achieving this would be to quantify over arbitrary propositions in each postcondition so that users could include a non-hierarchical proof fragment where necessary. To a certain extent, this is already supported because the $Data$ component of a postcondition is user-defined and therefore arbitrary. However, another option would be to add a separate conjunct to the postcondition.

\begin{figure}[ht]
\begin{centering}
\includegraphics[scale=0.4]{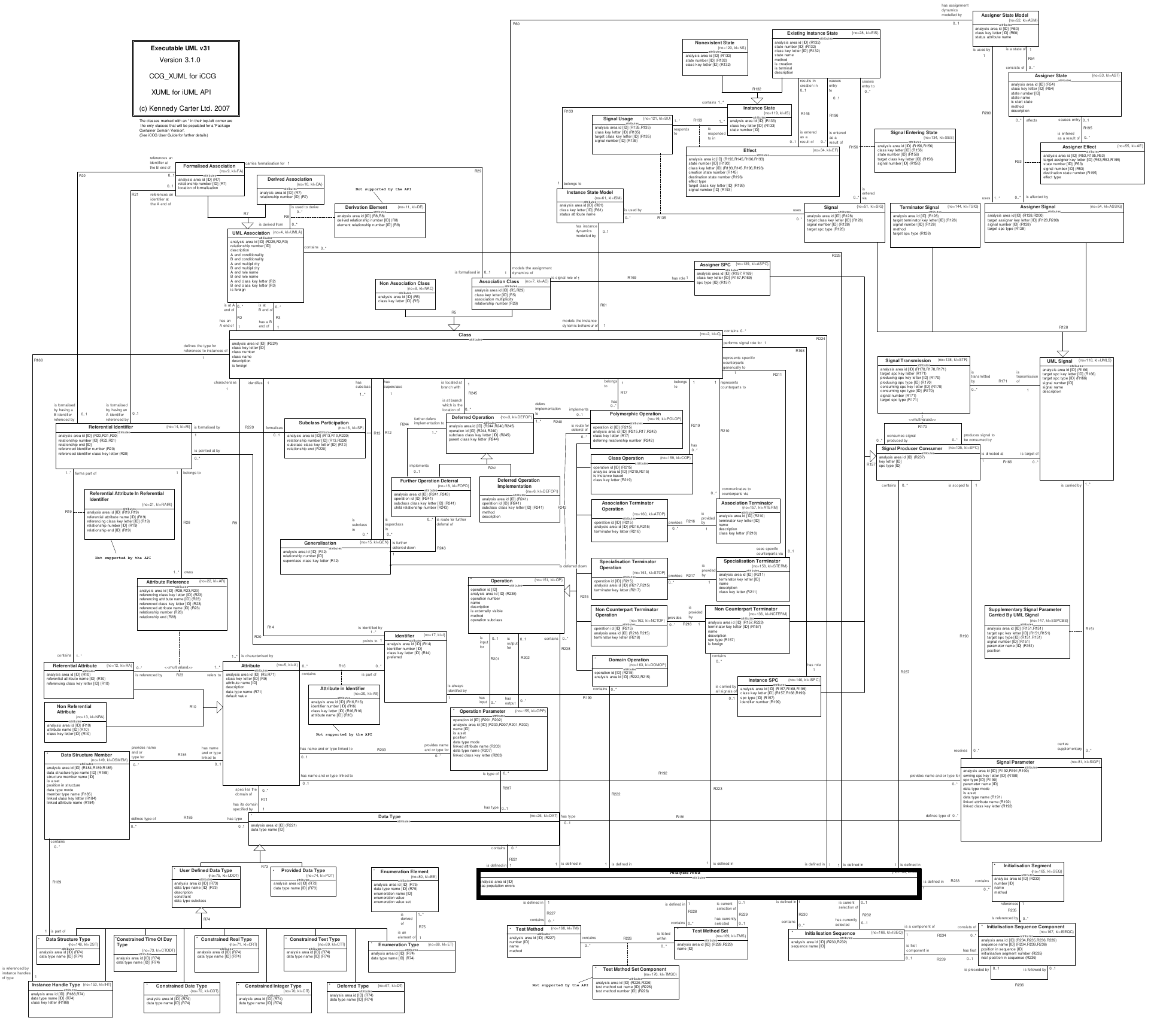}
\caption{An industrial strength model (of executable UML, courtesy of Abstract Solutions Ltd) comprising a strongly ordered core (admittedly, one defined by generalisation---an area for further study---as well as containment, and visually apparent only to a subject matter expert) surrounded by an unordered covering.}
\label{figure.a}
\end{centering}
\end{figure}

Clearly, we do not exclude the possibility of an ordered model being embedded within a larger model, like an ordered core with an unordered covering (for example, see Fig. \ref{figure.a}). In fact, our experience suggests that the majority of industrial models (which are characterised by their size rather than their complexity) are like this, and that the algorithms which transform them invariably perform preorder traversals of the ordered cores of the source models. That is not say that every model transformation fits this mould. However, it is reasonable to suppose that the lessons learnt from this study may also be applicable to other kinds of model transformations.

\section{Acknowledgements}

The authors would like to thank Iman Poernomo for his support in writing this paper.

\bibliographystyle{eptcs}
\bibliography{references}

\end{document}